\documentclass[apj,twocolumn]{aastex701}
\usepackage{amsmath}
\usepackage{CJK}
\usepackage{subcaption}
\usepackage{graphicx}
\usepackage{fmtcount}
\begin{document}
\begin{CJK*}{UTF8}{gbsn}

\title{Where Do Hot Jupiters Come From? Revisiting Tidal Disruption and Ejection in High-Eccentricity Migration}

\author[orcid=0009-0009-9688-9126]{Qianli Fan(范千里）}
\affiliation{School of Physics and Astronomy, Sun Yat-sen University, Zhuhai 519082, China}
\email{fanqli@mail2.sysu.edu.cn}

\author[orcid=0000-0002-9442-137X]{Shang-Fei Liu (刘尚飞)} 
\affiliation{School of Physics and Astronomy, Sun Yat-sen University, Zhuhai 519082, China}
\affiliation{CSST Science Center for the Guangdong-HongKong-Macau Great Bay Area, Sun Yat-sen University, Zhuhai 519082, China}
\email[show]{liushangfei@mail.sysu.edu.cn}
\correspondingauthor{Shang-Fei Liu}
 
\begin{abstract}

The origin of hot Jupiters remains a key open question. In the high-eccentricity migration scenario, traditional coreless models predict a strict tidal exclusion zone within $\sim 2.7$ tidal radii $r_\textrm{t}$, in which giant planets are either fully disrupted or ejected. We revisit this limit using three-dimensional hydrodynamic simulations of giant planets with realistic dense cores (10 - 20 $M_\oplus$). We find that even a few-percent-mass core fundamentally changes the outcome: \textbf{no total disruptions} occur within the previously suggested destruction zone ($\lesssim 2.7 \, r_\textrm{t}$). For deep encounters ($\lesssim 1.7 \, r_\textrm{t}$) planets suffer severe envelope stripping and are either progressively downsized to dense remnants or ejected after a few close encounters, possibly contributing to the free-floating planet population. In the intermediate regime ($  \sim 1.7  $--$2.0, r_\mathrm{t}$), planets experience significant partial mass loss over repeated encounters. For wider encounters ($  \gtrsim 2.0\, r_\mathrm{t}  $), mass loss is minimal, allowing the planets gradually circularize into hot Jupiters. Furthermore, we show that for highly eccentric orbits ($e\gtrsim 0.9$), the change in specific orbital energy $  \Delta E_{\mathrm{orb}}  $ depends primarily on periastron distance $  r_\mathrm{p}  $ rather than semi-major axis $  a  $. This enables us to extrapolate our fixed-$  a  $ results across a broad ($a$, $e$) parameter space and identify a well-defined tidal ejection zone whose sharp boundaries converge asymptotically. Our results highlight the crucial role of planetary internal structure in high-eccentricity migration and suggest that the survival and transformation of core-bearing giant planets are far more common than previously thought.
%Using coupled hydrodynamic and orbital integration schemes, we investigate the long-term tidal evolution of giant planets by simulating three successive encounters and extrapolating their outcomes. Our results suggest that planets are not merely destroyed or spared but may undergo divergent evolutionary pathways. Deep encounters can lead to tidal downsizing, potentially transforming Jupiter-mass planets into super-Earths, while a subset might be fully ejected, possibly contributing to the free-floating planet population. Wider encounters appear to result in circularization into hot Jupiters, with timescales constrained by the tidal quality factor $Q$. By mapping these outcomes across initial orbital parameters, we present a revised $a$--$e$ diagram that supersedes the classical picture, offering a unified framework for interpreting the origins of close-in planets.

\end{abstract}


\section{Introduction}
The discovery of hot Jupiters has posed a long-standing challenge to planet formation theory \citep{1995Natur.378..355M}. In the standard core accretion paradigm, giant planets are expected to form beyond the snow line, where solid material is abundant enough to assemble a multi-Earth-mass core before gas accretion can proceed \citep{1996Icar..124...62P}. This formation locus, at several astronomical units from the star, is incompatible with the observed orbits of hot Jupiters, necessitating a subsequent migration mechanism to bring them inward. Two broad classes of migration have been proposed to resolve this discrepancy: disk-driven migration \citep{1996Natur.380..606L} and high-eccentricity migration \citep{1996Sci...274..954R}.

In the disk-driven scenario, tidal interactions between the planet and the protoplanetary disk exchange angular momentum, causing the planet to spiral inward. While this mechanism is widely invoked to explain the population of close-in planets, it faces several well-known limitations. 
First, Type I migration for Earth to Neptune mass planets has historically been regarded as rapid, potentially delivering planets into the star before the disk disperses. This timescale problem requires fine-tuned disk properties or planet traps to circumvent \citep{1997Icar..126..261W, 2003ApJ...588..494M}. 
%However, more recent work has shown that mechanisms such as corotation torques and thermal effects can significantly slow, stall, or even reverse inward migration, thereby mitigating this timescale issue \citep{2006A&A...459L..17P, 2008ApJ...678..483B, 2012ARA&A..50..211K}.
Second, Type II migration for gap-opening giant planets is coupled to the viscous evolution of the disk, making the final orbital distance sensitive to uncertain disk parameters and potentially halting too far out to produce the shortest-period hot Jupiters \citep{1996Natur.380..606L, 2004ApJ...604..388I}. 
However, more recent work has shown that in realistic disks with complex structures, such as rims, rings, and gaps, disk migration can facilitate inward or outward migration under certain conditions, including concurrent accretion and thermodynamic effects \citep{chen_preservation_2020, li_3d_2023}. Also, mechanisms such as corotation torques and thermal effects can significantly slow, stall, or even reverse inward migration, thereby mitigating this timescale issue \citep{2006A&A...459L..17P, 2008ApJ...678..483B, 2012ARA&A..50..211K}. Third, and perhaps most critically, disk migration naturally produces nearly circular orbits, whereas a fraction of hot Jupiters with moderate eccentricities is the hallmark that tidal circularization is in effect \citep{2018ARA&A..56..175D}. 

High-eccentricity migration (HEM) provides an alternative pathway to form close-in gas giants. In this scenario, gravitational interactions among massive bodies can excite a giant planet onto a highly eccentric orbit. Several dynamical mechanisms have been identified as capable of producing such extreme eccentricities, including planet–planet scattering \citep{1996Sci...274..954R, 2008ApJ...686..621F}, Lidov-Kozai oscillations in binary or multi-body systems \citep{1962P&SS....9..719L, 1962AJ.....67..591K, 2007ApJ...669.1298F, 2011Natur.473..187N}, and secular chaos in closely packed multi-planet systems \citep{2011ApJ...735..109W, 2017ApJ...848...20W}.
%Some observational evidences lend support to this scenario. 
The eccentric orbits of many warm Jupiters and cold Jupiters are consistent with predictions from scattering and Kozai-Lidov dynamics \citep{2014Sci...346..212D, 2025ApJ...980L..31W}. 
Moreover, population level studies indicate that the observed properties of hot Jupiter systems, including the prevalence of outer giant companions and their orbital characteristics, strongly favor high eccentricity migration as the dominant formation mechanism\citep{2023ApJ...956L..29Z}.
Furthermore, the discovery of hot Jupiters with orbits that are misaligned or even retrograde relative to their host star's spin axis aligns naturally with predictions of high-eccentricity migration, whereas such configurations are difficult to explain via smooth disk migration alone \citep{2009ApJ...696.1230F, 2010ApJ...725.1995M}. In addition, recent demographics of hot Jupiters exhibit a stellar age dependency, suggesting a multichannel formation scenario wherein a substantial portion of the population arrives the vicinities of their host stars on a timescale of several Gyrs, thereby supporting HEM as a key late-forming pathway \citep{chen_evolution_2023, chen_origin_2025, schmidt_most_2026}.

Although the HEM scenario for hot Jupiters is increasingly accepted as essential for explaining certain population features, the detailed tidal evolution remains inadequately addressed. Tidal theory for close encounters was pioneered by \citet{1977ApJ...213..183P} and \citet{1981A&A....99..126H}, who implement impulsive approximations for tidal energy deposition and constant time lag models for energy dissipation and orbital circularization, respectively. %  initially developed in the context of stellar interactions \citep{1977ApJ...213..183P, 1981A&A....99..126H}, establishing the framework for understanding energy dissipation and orbital circularization. 
However, these models inadequately capture initial high-eccentricity states in HEM: the impulsive approximation fits single parabolic encounters, not repeated periastron passages in bound orbits, while the constant time lag assumes weak, linear tides that fail under extreme eccentricities where nonlinear effects prevail. \citet{carter_tidal_1983, carter_mechanics_1985} make use of a simplified affine star model to study tidal deformation, however, mass loss is explicitly excluded.

To overcome these limitations, particularly the potential for significant mass loss from the planet under intense tidal forces, hydrodynamic simulations, modeling giant planets initially at $\sim 5$ AU that scattered into highly eccentric orbits, have been employed to capture tidal stripping and energy dissipation during single and repeated close encounters with sun-like stars \citep{2005Icar..175..248F, 2011ApJ...732...74G, 2013ApJ...762...37L}. When characterizing the strength of the tidal field exerted on a planet by its host star, it is convenient to define a characteristic tidal radius as :
\begin{equation}
    r_{\mathrm{t}} = R_{\mathrm{P}} \left( \frac{M_{\star}}{M_{\mathrm{P}}} \right)^{\frac{1}{3}},
    \label{eq:tidal_radius}
\end{equation}
where $M_\star$, $M_\mathrm{P}$ and $R_\mathrm{P}$ are the stellar mass, planetary mass and radius, respectively. 

\citet{2005Icar..175..248F, 2011ApJ...732...74G} concur that, for single encounters giant planets survive tidal stripping and experience no mass loss at pericenter distances $r_\mathrm{p} \gtrsim 2.2 \, r_\mathrm{t}$, and tidally excited normal modes within the planet deposit orbital energy via adiabatic oscillations, ultimately tightening the orbit upon dissipation. For deeper passages ($r_\mathrm{p} \lesssim 1.8 \, r_\mathrm{t}$), intense tidal interactions induce modest to substantial envelope stripping, often exceeding $\sim$ 50\% mass loss in extreme cases. Moreover, the stronger tidal force on the planet's near side to the star causes asymmetric mass removal. Because the orbital energy is nearly uniformly distributed in the tidal debris, this asymmetry imparts a positive velocity kick to the remnant, increasing its specific orbital energy and often dominating the decrease caused by internal energy deposition from tidal oscillations, thereby resulting in ejection rather than orbital tightening in deepest encounters. 

\citet{2011ApJ...732...74G} extended this to simulate multiple periastron passages by reducing the semi-major axis to $\sim 0.1$ AU, shortening the orbital period by more than two orders of magnitude. They find that multiple passages result in progressive mass loss even for cases with no mass loss in their initial passages, and define an exclusion zone at $r_\mathrm{p} \leq 2.7 \, r_\mathrm{t}$, within which gas giants are either tidally destroyed or ejected. However, both \citet{2005Icar..175..248F} and \citet{2011ApJ...732...74G} adopted an $\Gamma = 2$ (or $n=1$) polytropic model, assuming a fully convective, coreless giant planet. A notable caveat of the $  n=1  $ polytrope is its fixed radius during adiabatic mass loss, which reduces density and explains why coreless giant planets are inevitably doomed to destruction in single or multiple passages once the cumulative mass loss becomes significant.

In contrast, \citet{2013ApJ...762...37L} demonstrated that the inclusion of a dense metallic core even as small as 10 $M_\oplus$ dramatically reverse these outcomes. Such a structure is expected from the core accretion paradigm \citep{1986Icar...67..391B, 1996Icar..124...62P}, while debate persisits on whether the metal distribution is highly concentrated or diffusive \citep{liu_formation_2019}. Nonetheless, stellar tides have minimal impact on a core with a central density $\sim 20 \,\mathrm{g}\,\mathrm{cm}^{-3}$. Using three-dimensional hydrodynamical simulations with composite polytropic models (e.g., a dense $  n=0.5  $ inner core surrounded by an $  n=1  $ envelope), \citet{2013ApJ...762...37L} showed that such cores allow planets to withstand periastron distances as close as $  \sim 1.2\,r_\mathrm{t}  $ with less mass loss compared to coreless scenarios. The core acts as a gravitational anchor, binding more of the gaseous envelope and reducing ejection likelihood. Consequently, the remnant experiences preferential envelope stripping, leading to a substantial metallicity enrichment and potential transformation into close-in super-Earths or hot Neptunes \citep{dong_lamost_2018, 2024AJ....168..132N}. %This mechanism provides a viable pathway for forming the observed population of dense, short-period rocky planets via HEM and partial tidal disruption, highlighting the critical role of internal structure in tidal evolution.

%of simplified polytropic models suggested that Jupiter-mass planets scattered into orbits with pericenter distances smaller than a critical value are highly likely to be either completely disrupted or ejected from the system \citep{2005Icar..175..248F, 2011ApJ...732...74G}, leaving a strict tidal exclusion zone where no bound remnant can survive. 
%However, these studies neglected a key feature expected from the core accretion paradigm \citep{1986Icar...67..391B, 1996Icar..124...62P}. The core, even if it contains only a few percent of the planet's mass, can fundamentally alter the tidal response by increasing the overall binding energy and providing a rigid inner boundary that suppresses large-scale envelope deformation \citep{2013ApJ...762...37L}. 

Since \citet{2013ApJ...762...37L} investigated only single pariastron passages, a key open question is whether core-baring giant planets can survive repeated passages without full disruption or ejection. Moreover, mass loss and thus the evolution of the specific orbital energy is highly sensitive to the planetary inerior structure. Revisiting the origins and initial orbital conditions of hot Jupiters formed via the HEM pathway is therefore essential, as comparing these with their formation sites \citep{1981PThPS..70...35H, 1988Icar...75..146S, 2011ApJ...743L..16O} could elucidate the mechanisms exciting high eccentricities.

%Moreover, most previous simulations have been limited to single pericenter passages, with subsequent evolution extrapolated using simplified prescriptions that cannot capture the complex coupling between mass loss, structural rearrangement, and orbital dynamics over multiple encounters. For planets on longer period orbits (say $a \gtrsim 1$ AU), simulating consecutive close encounters with full hydrodynamics remains computationally challenging. Studies that have attempted multiple passages often circumvent this difficulty by adopting artificially short orbital periods (e.g., \cite{2011ApJ...732...74G}) to reduce computational expense. While computationally expedient, this approach yields orbital periods substantially shorter than those typical of planets undergoing high-eccentricity migration from their formation sites beyond the snow line \citep{1981PThPS..70...35H, 1988Icar...75..146S, 2011ApJ...743L..16O}.

%In this paper, we revisit multiple tidal encounters using three-dimensional hydrodynamic simulations of giant planets with dense cores, combining full hydrodynamics with approximate spherical reconstruction between passages. For highly eccentric orbits ($e\gtrsim0.9$), $\Delta E_{\mathrm{orb}}$ depends primarily on $r_{\mathrm{p}}$, allowing extrapolation to the full $a$ - $e$ space and delineating where hot Jupiters can form. 
The paper is organized as follows. In Section \ref{sec:methods}, we describe our numerical methods. The remnant mass and orbital energy evolution obtained from our hydrodynamic simulations is presented in Section \ref{sec:results}. We discuss implications, comparison with coreless models, and generalization to the $a$ -- $e$ plane in Section \ref{sec:discussion}. We summarize and outline future directions in Section \ref{sec:summary}.

\section{Methods}
\label{sec:methods}
\subsection{Setup of Hydrodynamic Simulations}
\label{sec:hydro}
%Our goal is to efficiently simulate the long-term tidal evolution of a giant planet from a high-eccentricity orbit. 
To investigate the response of the giant planet during close encounters with its host star, we perform three-dimensional hydrodynamic simulations using \textsc{FLASH}, an adaptive-mesh, grid-based hydrodynamics code \citep{2000ApJS..131..273F}.
%The star is treated as a point mass with $M_{\star} = 1\,M_{\odot}$, as energy dissipation in the star is negligible compared to that in the planet for Jupiter-like planets and solar-type hosts \citep{2005Icar..175..248F}. 
The star is treated as a point mass with $M_{\star} = 1\,M_{\odot}$. This point-mass approximation for the star is justified by the scaling relation of \cite{2005Icar..175..248F}, who showed that the ratio of tidal energy dissipated in the planet to that in the star scales as 
\begin{equation}
    \frac{\Delta E_\mathrm{P}}{\Delta E_{\star}}\simeq\left(\frac{M_{\star}}{M_\mathrm{P}}\right)^2\left(\frac{R_\mathrm{P}}{R_{\star}}\right)^5.
\end{equation}
For a solar-type star and a Jupiter-mass planet, this ratio is $\sim 10$, indicating that the planet's tidal response dominates the interaction. Consequently, the back-reaction on the stellar structure and its contribution to orbital evolution are subdominant for the systems considered here. Nevertheless, we caution that this assumption may break down for lower-mass, fully convective stars, or during later stages of stellar evolution, where stellar tides could become significant. The present simulations therefore apply to interactions between giant planets and main-sequence, solar-type stars.
For the hydrodynamic integration, we employ the directionally split piecewise parabolic method \citep{1984JCoPh..54..174C} within the FLASH framework. The gravitational treatment follows the modified gravity algorithm developed by \cite{2011ApJ...732...74G}, and we configure the multipole gravity solver with the same parameter choices as \cite{2013ApJ...767...25G}. All simulations are carried out in the rest frame of the planet. 

%The planet is modeled with composite polytropes following the treatment in \citep{2013ApJ...762...37L}.
The planet is modeled as a two-layer composite polytrope following the treatment of \cite{2013ApJ...762...37L}. This representation consists of a dense, stiff core (with polytropic index $n_{1}=0.5$) surrounded by a gaseous envelope (polytropic index $n_{2}=1$), capturing the distinct equations of state between the core and envelope. Our detailed setup is provided in the \hyperref[app:profiles]{Appendix}. The simulation domain is a cube with side length $2\times10^{13}$ cm, initialized with a single root block of $8^3$ cells. We employ the PARAMESH library \citep{1987CMAME..61..323L} to allow up to 16 levels of adaptive mesh refinement based on local flow conditions, resulting in a minimum grid size of approximately $0.01\,R_{\mathrm{P}}$. 
For all simulations, the planet is initialized on an orbit with a fixed semi-major axis of $1$ AU, and is initially placed at a separation of $3\,r_{\mathrm{t}}$ from the star, set to be non-rotating. This distance is sufficient to ensure that tidal forces are weak and do not significantly perturb the orbit over the short initial segment of the trajectory simulated here.
%\begin{equation}
 %   r_{\mathrm{t}} = R_{\mathrm{P}} \left( \frac{M_{\star}}{M_{\mathrm{P}}} \right)^{\frac{1}{3}},
%    \label{eq:tidal_radius}
%\end{equation}
%where $M_{\mathrm{P}} = 1\,M_{\mathrm{J}}$ and $R_{\mathrm{P}} = 1\,R_{\mathrm{J}}$ are the initial planetary mass and radius. 
For our adopted parameters, $r_{\mathrm{t}} \simeq 0.00475$ AU, slightly larger than the solar radius ($0.00465$ AU).

%For a given pericenter distance $r_{\mathrm{p}}$, we perform hydrodynamical simulations to model the planet's response during a single close passage by its host star (treated as a point mass). Once the planet undergoes a close encounter without being ejected or completely disrupted, the remnant will subsequently settle into a Keplerian orbit. The orbit is determined by hydrodynamic simulation. This orbital state can be efficiently integrated forward in time, with the resulting parameters serving as initial conditions for subsequent encounters. We now turn to the task of defining the planetary structure both before and after tidal encounters.

\subsection{An Approximate Scheme for Repeating Encounters}
\label{sec:scheme}
Our goal is to understand the long-term tidal evolution of a giant planet on a high-eccentricity orbit. However, simulating many consecutive close encounters with full hydrodynamics for orbits with large initial semi-major axes is computationally prohibitive. To overcome this limitation, we employ a hybrid approach that combines full hydrodynamic simulations of individual strong encounters with a simplified treatment of the orbital and structural evolution between encounters.

For a given pericenter distance $r_{\mathrm{p}}$, we first perform a hydrodynamic simulation of a single close passage. Once the planet undergoes a close encounter without being ejected or completely disrupted, the remnant will subsequently settle into a Keplerian orbit. This orbital state can be efficiently integrated forward in time, allowing us to readily determine the orbital parameters for the remnant's subsequent pericenter passage.
We terminate the simulation when the remnant reaches a distance exceeding $  20\, r_\mathrm{t}  $ from the host star and its orbital energy has stabilized, at which point it has entered a stable Keplerian orbit. The resulting remnant is non-spherically symmetric.

Since our simulations are frictionless, the remnant does not relax back to a spherical configuration within the hydrodynamical integration. However, the oscillation modes excited during the close encounter are expected to damp on a timescale that can be estimated using the tidal quality factor $Q$. For a weakly damped harmonic oscillator, the amplitude decays as 
\begin{equation}
    A(t) = A_0 \exp(-\pi t/\tau_{\text{damp}}),
\end{equation}
where the damping timescale \(\tau_{\text{damp}}\) is related to the oscillation period \(P_{\text{osc}}\) and \(Q\) by

\begin{equation}
\tau_{\text{damp}} = \frac{Q}{\pi} P_{\text{osc}}.
\end{equation}
At \(t = \tau_{\text{damp}}\), the amplitude is reduced to \(e^{-\pi} \approx 4\%\) of its initial value.
The \(f\)-mode period scales as \(P \sim 2\pi/\sqrt{G\bar{\rho}}\) \citep{1989nos..book.....U}, which for Jupiter's mean density \(\bar{\rho} \approx 1.33\)~g~cm\(^{-3}\) yields \(P \sim 2.5\)~hours.
Moreover, \cite{1999P&SS...47.1211G} computed the free oscillation spectrum of Jupiter and found acoustic mode frequencies of approximately 152--155~\(\mu\)Hz, corresponding to periods of about 1.8--1.9~hours for low-degree modes. 
Adopting \(Q \sim 10^{4}\) \citep{2009Natur.459..957L}, the corresponding damping timescale \(\tau_{\text{damp}}\) is estimated to be approximately $0.73$ years. For a more conservative value appropriate for hot Jupiters, \(Q \sim 10^{6}\), \(\tau_{\rm damp}\) increases to $73$ years (corresponding to $  a\approx 17.5  $ AU). We note that this estimate is based on linear theory and therefore represents a conservative upper limit. In the highly dynamical regime of our simulations, which involve large-amplitude oscillations and substantial mass loss, nonlinear effects and energy/momentum removal by the ejected material are expected to damp the oscillations more rapidly. Consequently, the assumption that the planet relaxes back to a nearly spherical configuration between encounters remains reasonable for a substantial and astrophysically relevant portion of the parameter space.
%Given that our simulations adopt an initial orbital period of one year, and noting that realistic orbital periods in high eccentricity migration are typically a decade or longer, the assumption of spherical relaxation between encounters is reasonable for a substantial portion of the parameter space.
%Given that our simulations adopt an initial orbital period of one year, we can reasonably assume that the planet returns to a spherically symmetric state before each subsequent encounter.
Therefore, we can attempt to construct a new, spherically symmetric planetary model to replace this remnant for the next encounter. 

%It should be noted that rotation is neglected in our simulations. The reconstruction approach will be detailed in the following section.

It should be noted that rotation is neglected in the present simulations. Because the initial spin orientation of the planet is expected to be random with respect to the orbital plane, and because the phase of tidally excited oscillations at each periastron passage could also be effectively random, including spin would introduce orientation-dependent tidal torques and potentially lead to chaotic evolutionary behavior over repeated encounters \citep{2011ApJ...732...74G}. To isolate the intrinsic per-encounter response of the planet, we deliberately reconstruct the remnant into a non-rotating, spherically symmetric configuration before each subsequent passage. Nevertheless, to address the possible importance of spin evolution, we have computed the net change in angular momentum of the bound remnant after each encounter (see Section \ref{sec:results}). These calculations show that the accumulated spin angular momentum remains well below the break-up limit and does not significantly alter our main conclusions.

%We then determine whether the material in each grid is bound to the planet by calculating its specific binding energy. Thus, we can obtain the total bound mass.
% To maintain a constant planetary radius in our model, we adjust the ratio of mean molecular weight in the core and the envelope. 
%\section{Hydrodynamical simulations of repeating tidal encounters}

\subsection{Planetary Structure and Post-Encounter Relaxation}
\label{sec:structure}

%In this work, we conducted simulations across two different core masses, with the initial pericenter distance $r_{\mathrm{p}}$ covering a range from 1.2 to 2.0 $r_{\mathrm{t}}$. All initial orbits of the planets are set to have a simi-major-axis of 1 AU. Each configuration was evolved for three orbital periods.
We model the giant planet as a two-layer composite polytrope containing a dense core, following the approach of \cite{2013ApJ...762...37L}. These polytropic equilibrium models follow the formalisms comprehensively established in \cite{2004ASSL..306.....H}. This structure is a more realistic representation than a single polytrope, as it accounts for the distinct mean molecular weight and equation of state between a silicate/metallic core and a gaseous envelope. The presence and mass of this core are the key parameters that differentiate our models from prior studies of tidal disruption.
The planets are modeled using the gamma-law equation of state
\begin{equation}
    e_i = \frac{P}{\rho(\gamma - 1)},
\end{equation}
where $e_i$ is the specific internal energy, and the gas is assumed to be ideal, calorically perfect, and adiabatic. 
The polytropic exponents are set to $\gamma_1 = 3$ in the core and $\gamma_2 = 2$ in the envelope. Here we equate the polytropic indices with the adiabatic indices, an assumption valid for Jupiter-like planets. 
The initial planetary radius $R_\mathrm{P}$ is 1 $R_\mathrm{J}$ and mass $M_\mathrm{P}$ is equal to 1 $M_\mathrm{J}$. 

If a planet loses mass on a timescale between its dynamical time and Kelvin-Helmholtz time, its structure evolves adiabatically, and the entropy profile is approximately preserved \cite{2013MNRAS.434.2940D}. As established by \cite{1987ApJ...318..794H}, under the relation $\gamma = 1 + 1/n$, the radius of a single-layered polytrope after mass loss for the specific case of $n=1$ remain constant. In the case of our two-layer composite polytrope, the presence of the core causes the radius to undergo a slight contraction. \cite{2013ApJ...762...37L} calculated the adiabatic response curves for varying degrees of mass loss for a composite polytrope with indices $n_1 = 0.01$ and $n_2 = 1$. Their results demonstrated that even with mass loss up to 80\%, the radii for models with 10 and 20 earth mass cores contracted to only about 90\% and 85\% of their original sizes, respectively. Based on these considerations, we therefore adopt the approximation that the planetary radius remains constant between encounters. 
In reality, the radius evolution is dominated by adiabatic contraction associated with mass loss. Two secondary competing effects, i.e., tidal heating during close periastron passages and subsequent radiative cooling, are expected to affect the final radius evolution. Consequently, the combined effect of these processes on the actual radius remains uncertain and is left for future self-consistent modeling. Our fixed radius assumption therefore serves as a reasonable simplified baseline for the three-consecutive-encounter simulations presented in this study.
Concurrently, we assume the presence of a rigid, stiff core. 
We initialize each simulation by fixing the core mass $M_\mathrm{core}$, core radius $R_\mathrm{core}$, $\gamma_{1}$, $\gamma_{2}$ and the total planetary radius $R_\mathrm{P}$, to approximate the adiabatic structural response.  The resulting density and mass profiles for different mass-loss scenarios are shown and discussed in \hyperref[app:profiles]{Appendix}.
%Figure \ref{fig:profiles} shows the density and mass profiles for models (core mass=$20 M_\oplus$) with three different mass-loss scenarios: no mass loss, 50\% mass loss, and 80\% mass loss. Although the central density of the core differs among models of varying total mass due to the nature of the polytropic model, they share a fixed core boundary radius and core mass. This consistency aligns with our rigid core assumption.

\section{Results}
\label{sec:results}

In this work, we follow the setup of the interior structure implemented in \citep{2013ApJ...762...37L} and simulate repeated tidal disruptions for two different core mass planets on an eccentric orbit with a semi-major axis of 1 AU, exploring initial pericenter distance $r_\mathrm{p}$ in the range of 1.2 -- 2.0 $r_\mathrm{t}$. Each model was followed through three successive periastron passages. The choice of $  a = 1  $ AU is deliberately much smaller than the single passages in previous studies \citep{2005Icar..175..248F, 2011ApJ...732...74G, 2013ApJ...762...37L}, which typically adopted $  a \sim 2.5  $ AU. This reduced semi-major axis ensures that even substantial positive velocity kicks from asymmetric mass loss during deep encounters remain insufficient to unbound the planet after the first passage, thereby allowing us to reliably track cumulative mass loss and remnant evolution over multiple tidal interactions. The evolution of the bound remnant mass after each encounter across all simulated cases is shown in the Figure \ref{fig:mass_loss}.

To validate the assumption of a fixed planetary radius between encounters, we show the most extreme mass-loss event ($r_\mathrm{p}/r_\mathrm{t} = 1.2$ and $M_\mathrm{core} = 10\,M_\oplus$) in our simulations in Figure \ref{fig:snapshots_profile}. The upper panels illustrate the planet's evolution across three periastron passages, including the reconstructed, fixed-radius remnant used as the initial condition for the subsequent encounter. The lower panels compare the spherically averaged density profiles and enclosed-mass profiles of the actual post-encounter remnant with the reconstructed model after the first and second passages, respectively. 

We determine whether material in each grid cell is bound to the planet by computing its specific binding energy relative to the remnant center; the remnant mass is then obtained by summing the masses of all bound cells. The resulting asymmetric post-encounter remnant is reconstructed into a simplified, spherically symmetric model to initialize the next simulation. The reconstructed profile provides a good match to the post-encounter structure over a large radial extent, with more than 75\% of the remnant mass enclosed within $1\,R_\mathrm{J}$.

To further assess the spin state of the remnant, we computed its total angular momentum after each encounter and compared it to the theoretical break-up angular momentum of a uniform spherical body with the same total mass and radius. For the deepest encounter simulated ($  r_p/r_t = 1.2  $, $  M_\mathrm{core} = 10\,M_\oplus  $), the remnant acquires approximately 41\% of the break-up value after the first passage. This fraction decreases to 33\% after the second passage and 27\% after the third. This declining trend reflects the weakening of tidal torques as the planet loses envelope mass and becomes more centrally condensed. Importantly, the accumulated angular momentum remains well below the break-up limit throughout, indicating that spin-up is self-limiting and does not threaten the structural stability of the remnant. Because the phase of the tidally excited bulge relative to the star-planet line at periastron is effectively random between successive encounters, each close passage can produce either a tidal spin-up or a tidal spin-down. Consequently, the net angular momentum of the remnant does not increase indefinitely but instead tends to saturate over many encounters.

%As a further check on the spin state of the remnant, we estimate the post-encounter angular momentum and compare it to the theoretical break-up angular momentum for a uniform sphere of the same mass and radius. For the deepest simulated encounter with a $10\,M_{\oplus}$ core, the angular momentum after the first passage is 41\% of the break-up value, decreasing to 33\% after the second and 27\% after the third. This trend reflects the weakening of tidal torques as the planet becomes more centrally condensed, and it indicates that spin-up remains self-limiting and well below the stability threshold.
%The upper panel illustrate the consequence of three periastron distance passages. compare the spherically-averaged density profile after the first encounter with the density profile used as the initial condition for the second encounter. 
% , showing a direct comparison before and after the encounter for the case of . The upper panel shows snapshots before and after encounters, while the lower panel compares the corresponding density and mass profiles. 
%(We determine whether the material in each grid cell is bound to the planet by calculating its specific binding energy. The masses of all bound cells are then summed to obtain the remnant mass.)
%The asymmetric post-encounter remnant is reconstructed into a simplified, spherically symmetric model to serve as the initial condition for the next simulation. The profile shows a good fit over a large fraction of the entire radial extent, over 75\% of the remnant's mass is enclosed within 1 $R_\mathrm{J}$. 

\subsection{The State of the Post-Encounter Remnant}
\begin{figure}[h]
    \centering
    \includegraphics[width=1\columnwidth]{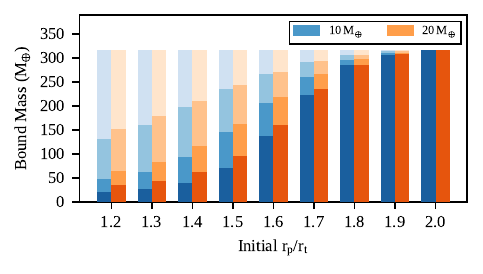}  
    \caption{Bound mass versus initial $r_\mathrm{p}/r_\mathrm{t}$ for planets with $10\,M_\oplus$ (blue) and $20\,M_\oplus$ (orange) cores. Progressively darker shades within each color group represent the initial state and the remnant mass after the first, second, and third encounters, respectively. All masses are in Earth masses.  } 
    \label{fig:mass_loss}
\end{figure}

\begin{figure*}[h]
    \centering
    \includegraphics[width=0.95\linewidth]{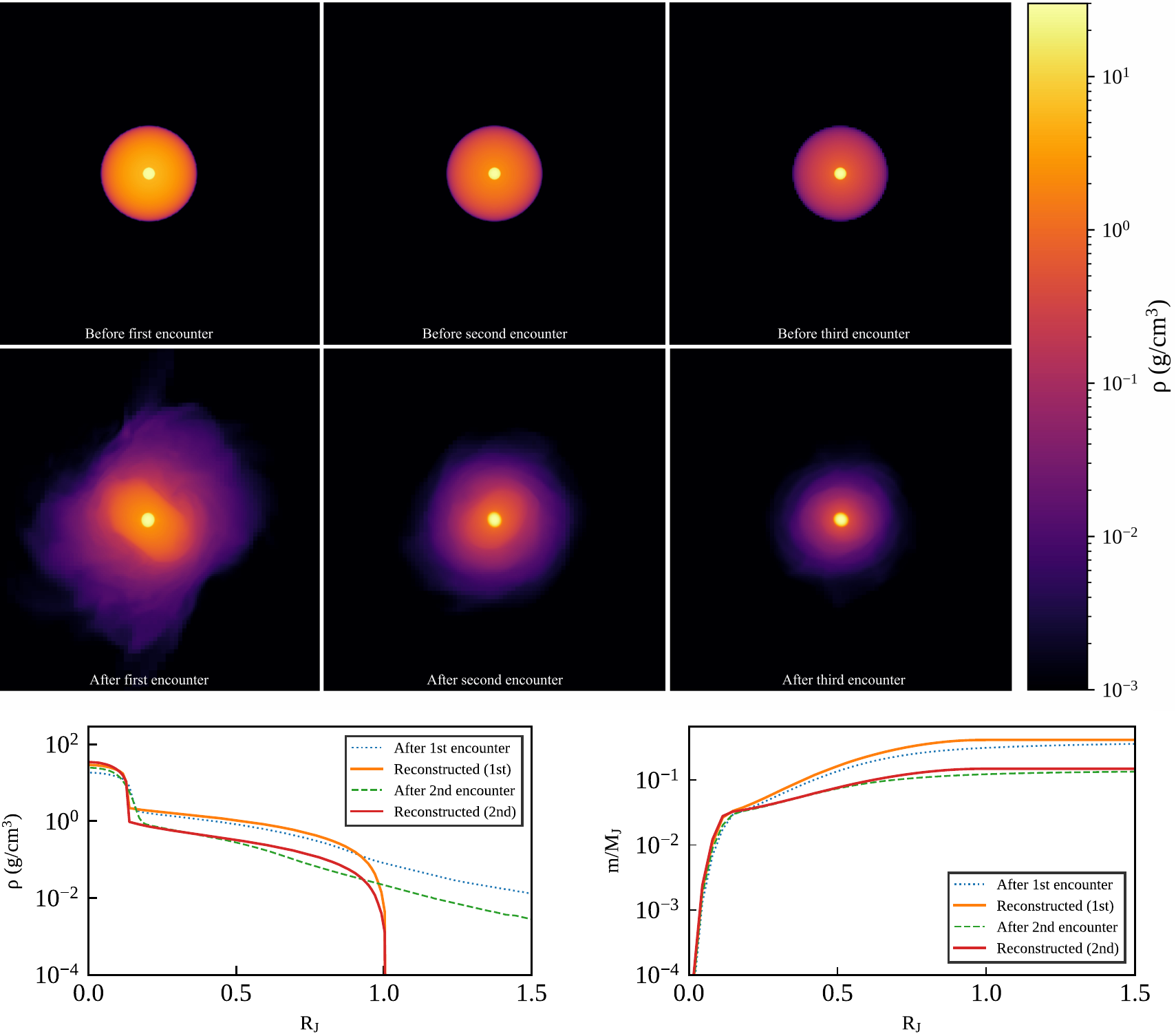} 
        
    \caption{Structural evolution during the most extreme mass-loss case in our simulations ($r_\mathrm{p}/r_\mathrm{t} = 1.2$, $M_{\mathrm{core}} = 10\,M_\oplus$). Upper panel: Density slices showing the planet before and after tidal encounters. The post-encounter remnant is visibly distorted and spinning, with a significant portion of the envelope stripped. Lower left: Quantitative comparison of the spherically averaged density profiles before (solid line) and after (dot and dashed line) the encounter, which provides a good fit over most of the radial extent. 
    Lower right: Cumulative mass profile comparison, showing that over 75\% of the remnant's mass is enclosed within $1\,R_\mathrm{J}$. The good agreement between the pre-encounter and post-encounter profiles over most of the radial extent justifies our approach of treating each encounter independently with the remnant relaxed back to a spherical configuration.}
    \label{fig:snapshots_profile}
\end{figure*}

During each close encounter, tidal stripping removes material from the planet's outer layers. As shown in Figure \ref{fig:snapshots_profile}, the remaining remnant is left spinning and adiabatically expanded, acquiring a structure distinct from that of an unperturbed planet.
We now characterize the physical state of the remnant after the first, second, and third tidal encounters. Our analysis focuses on two key properties: the remaining bound mass and the morphological structure.
For planets with a dense core, mass loss increases significantly as the initial $r_\mathrm{p}/r_\mathrm{t}$ decreases. This effect is further amplified with each successive encounter. 
Figure \ref{fig:mass_loss} compares two planetary masses ($10\,M_\oplus$ in blue and $20\,M_\oplus$ in orange) across nine initial $r_\mathrm{p}/r_\mathrm{t}$ ratios. Within each mass group, progressively darker shades represent the bound mass after the initial state and after the first, second, and third encounters (in $M_\oplus$). 

At these deep encounters ($r_\mathrm{p}/r_\mathrm{t}\lesssim1.4$), tidal stripping is catastrophic. After just one passage, the planet loses the majority of its gaseous envelope; by the third encounter, both the $10\,M_\oplus$ and $20\,M_\oplus$ core models are stripped down to little more than their bare cores, retaining only a thin residual envelope comparable in mass to the core itself (Figure \ref{fig:mass_loss}). The remnant consists essentially of the dense core surrounded by a tenuous, loosely bound gas layer that is highly susceptible to further erosion by photoevaporation or thermal escape. These objects represent a potential pathway to the formation of bare cores.

The regime of intermediate separations centers around $r_\mathrm{p}/r_\mathrm{t}\sim1.6$. In this regime, mass loss is substantial but not complete. The remnant retains a significant fraction of its envelope, but its structure is modified: the remaining gas is expanded and more weakly bound, making it vulnerable to continued stripping in subsequent encounters. This is also the regime where orbital energy changes are most pronounced (see section \ref{sec:Orbital Evolution}), with some planets experiencing net energy gain that can lead to ejection after multiple passages.

At these larger pericenter distances ($r_\mathrm{p}/r_\mathrm{t}\sim2.0$), tidal effects are mild. A single encounter removes only a small fraction of the envelope mass, and the cumulative effect after three encounters remains negligible. The planet's structure is almost unchanged. It retains its original radius and nearly all of its envelope mass. These planets are the most likely to survive multiple passages and eventually circularize into hot Jupiters over Gyr timescales.

It is instructive to contrast our findings with those of \cite{2011ApJ...732...74G}, who simulated tidal encounters of coreless giant planets under similar pericenter conditions. In their models, every planet with \(r_{\mathrm{p}}/r_{\mathrm{t}} \leq 2.0\) was either completely disrupted or ejected within at most three encounters. This stands in stark contrast to our results, where even at the closest separations we find no total disruptions; instead, planets are progressively stripped down to a bare core that remains gravitationally bound and structurally intact. The fundamental difference lies in the presence of a dense, incompressible core in our models. A detailed comparison of the underlying physical mechanisms and a discussion of why our orbital setup leads to deterministic rather than chaotic outcomes are presented in section \ref{sec:survival-and-orbit}.
%At wide separations ($r_\mathrm{p}/r_\mathrm{t} \sim 1.9-2.0$), a single encounter has only a minor impact on the structure, so the cumulative mass loss remains small and nearly identical between successive passes. 
%In contrast, at close separations ($r_\mathrm{p}/r_\mathrm{t} \sim 1.2-1.3$), both mass groups are stripped down to little more than their cores after three encounters, leaving a remnant envelope comparable in mass to the core itself. Overall, the $20\,M_\oplus$ planets retain a larger fraction of their initial mass than their $10\,M_\oplus$ counterparts, consistent with the findings of \cite{2013ApJ...762...37L}. As mass loss becomes more severe, the advantage of the higher core mass planet becomes increasingly evident. With the total mass significantly reduced, its initially larger core accounts for a greater fraction of the remnant.

\begin{figure*}[t] 
    \centering
    \includegraphics[width=\linewidth]{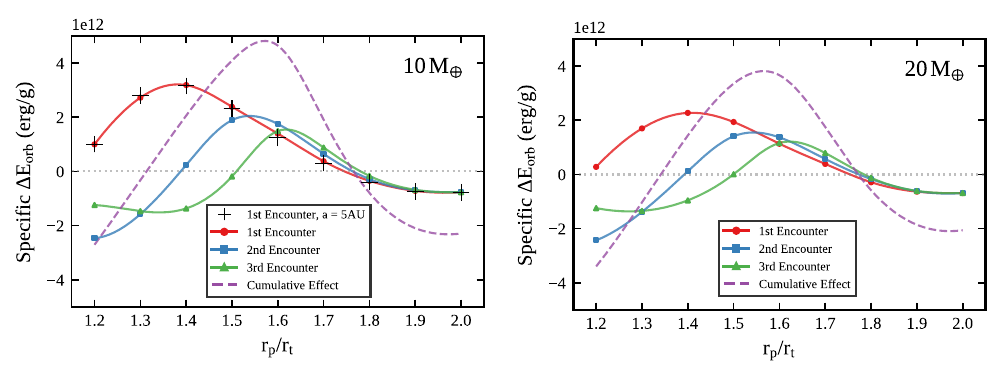} 
    \caption{Specific orbital energy change $\Delta E_{\mathrm{orb}}$ as a function of normalized pericenter distance $r_\mathrm{p}/r_\mathrm{t}$ for planets with $10\,M_\oplus$ (left) and $20\,M_\oplus$ (right) cores ($a = 1 $ AU). Colored symbols show the energy change after the first (red circles), second (blue squares), and third (green triangles) encounters. The discrete data points are connected by cubic spline interpolations to illustrate the trend. The horizontal dashed line marks $\Delta E_{\mathrm{orb}} = 0$, separating orbital energy gain (above) from loss (below). For both core masses, the peak of positive energy shifts toward larger $r_\mathrm{p}/r_\mathrm{t}$ with successive encounters, and the three curves converge as $r_\mathrm{p}/r_\mathrm{t} \rightarrow 2.0$, where mass loss becomes negligible. The black cross in the left panel shows a validation run with $a = 5$ AU (all other parameters fixed), confirming that $\Delta E_{\mathrm{orb}}$ is insensitive to semi-major axis for highly eccentric orbits.}
    \label{fig:orbital energy}
\end{figure*}

\begin{figure*}[t] 
    \centering
    \includegraphics[width=\textwidth]{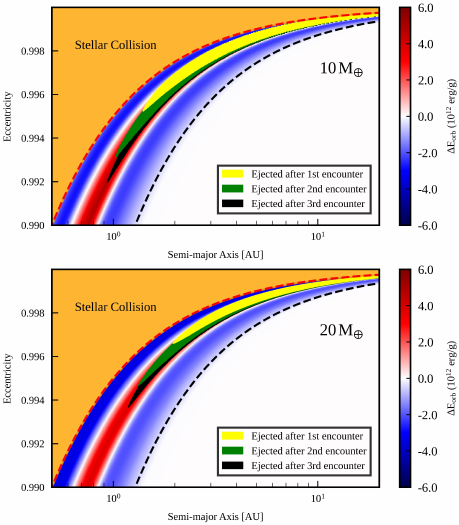} 
    \caption{Predicted Orbital Energy Change in the $\mathbf{a}-\mathbf{e}$ Parameter Space, $r_\mathrm{p}$ increases from the top-left to the bottom-right corner. The panels map the cumulative specific orbital energy change, $\Sigma\Delta E_{\mathrm{orb}}$, after three tidal encounters as a function of the initial orbital semi-major axis $\mathbf{a}$ and eccentricity $\mathbf{e}$, for planetary models with a $10\,M_\oplus$ (upper panel) and a $20\,M_\oplus$ (lower panel) core. The color indicates the net energy change (red: increase, blue: decrease). The yellow, green, and black shaded regions mark orbits that would be ejected after one, two, and three encounters, respectively. three encounters, respectively. The red dashed line denotes the stellar collision boundary($r_\mathrm{p} = 0.005AU$). The black dashed line at $r_\mathrm{p}/r_\mathrm{t} = 2.7$ indicates the critical disruption zone identified in previous coreless models.}
    \label{fig:ae_diagram}
\end{figure*}

\subsection{Orbital Evolution During Repeated Encounters}

\label{sec:Orbital Evolution}

The change in specific orbital energy, $\Delta E_{\mathrm{orb}}$, occurs predominantly during the brief but intense periastron passage.
Previous studies of single tidal encounters have shown that planets on wider initial orbits (larger semi-major axes) are prone to ejection following a single deep encounter with the host star \citep[e.g.,][]{2005Icar..175..248F, 2011ApJ...732...74G}. In contrast, \citep{2013ApJ...762...37L} demonstrate that the dominant factor determining the sign and magnitude of $  \Delta E_{\mathrm{orb}}  $ is asymmetric mass loss, which is highly sensitive to the interior structure (in particular, the presence and mass of a dense core). 

Here we compute $\Delta E_{\mathrm{orb}}$ across three consecutive periastron passages for giant planets with dense cores at various initial periastron distances. As described earlier, we adopt a reduced semi-major axis of $a = 1$ AU to prevent early ejection following deep passages and thereby enable reliable tracking of cumulative orbital evolution over multiple encounters. The resulting $\Delta E_{\mathrm{orb}}$ for individual passages as well as the cumulative effect after three passages are presented in Figure \ref{fig:orbital energy}. We have applied a cubic spline interpolation to smooth the data (curves in Figure \ref{fig:orbital energy}).

We observe qualitatively similar changes in specific orbital energy, $\Delta E_{\mathrm{orb}}$, after the first periastron passage to those reported by \citet{2013ApJ...762...37L}, despite our adoption of a substantially smaller semi-major axis ($  a = 1  $ AU). For highly eccentric orbits ($e \gtrsim 0.9 $), $\Delta E_{\mathrm{orb}}$ is predominantly determined by the periastron distance $r_\mathrm{p}$, exhibiting only weak dependence on the semi-major axis $a$. This behavior stems from the fact that, in the high eccentricity limit, the trajectories near pericenter become nearly identical for a fixed $r_\mathrm{p}$ irrespective of $a$. To test this, we performed an additional set of simulations with a fixed semi-major axis of $a=5\,\mathrm{AU}$ covering the same range of pericenter distances. The resulting $\Delta E_{\mathrm{orb}}$ values are shown as black crosses in Figure \ref{fig:orbital energy}. The differences relative to the $a=1\,\mathrm{AU}$ results are negligible (typically within 3.5\%). 

A key trend evident across all simulations is that planets with a more massive core ($20\,M_\oplus$) exhibit systematically smaller orbital energy changes than their $10\,M_\oplus$ counterparts. This difference arises from asymmetric mass loss and tidal energy dissipation. 
The orbital energy change in a tidal encounter is governed by two competing processes. First, tidal excitation transfers orbital energy into internal oscillation modes of the planet, which tends to decrease the orbital energy (making the orbit more bound). 
Second, asymmetric mass loss during the encounter can increase the orbital energy of the surviving remnant. When mass is stripped during a close encounter, the loss is not symmetric: the hemisphere facing the star experiences stronger tidal forcing and always loses more mass than the trailing hemisphere. This asymmetric mass loss acts as a momentum recoil. 
%The ejected material carries away an angular momentum that differs from that of the remnant, effectively imparting a net velocity kick to the surviving remnant. 
Because the mass lost from the facing side exceeds that from the trailing side, this recoil consistently increases the orbital energy of the remnant, making the orbit less bound. This asymmetry-driven energy gain is the primary mechanism that can lead to ejection after multiple encounters.
\cite{2013ApJ...762...37L} quantified this effect and found a linear correlation between the orbital energy change $\Delta E_{\mathrm{orb}}$ and the asymmetry of mass loss $\Delta m_2-\Delta m_1$, where $\Delta m_2$ and $\Delta m_1$ are the fraction of mass lost through $L_2$ and $L_1$, respectively. When the two tails are nearly equal in mass ($\Delta m_2 - \Delta m_1 \approx 0$), the asymmetry term is small and the orbital energy change is dominated by tidal dissipation, leading to a net decrease in orbital energy. When the mass loss is highly asymmetric, the remnant can receive a net energy gain. A detailed derivation and fitting of this linear relationship can be found in \citep{2013ApJ...762...37L}.
The stronger gravitational binding of the $20\,M_\oplus$ core suppresses both the total mass loss and the degree of asymmetry. A more massive core acts as an anchor, holding the envelope more tightly and reducing the deformation that leads to asymmetric stripping. 

%Examining the dependence on orbital scale, we find that for highly eccentric orbits,  the specific orbital energy change $\Delta E_{\mathrm{orb}}$ is predominantly a function of the pericenter distance $r_\mathrm{p}$, with only a weak explicit dependence on the semi-major axis $a$. This behaviour arises because, for highly eccentric orbits, the trajectories near pericenter are nearly identical for a given $r_\mathrm{p}$, regardless of $a$. To test this, we performed an additional set of simulations with a fixed semi-major axis of $a=5\,AU$ covering the same range of pericenter distances. The resulting $\Delta E_{\mathrm{orb}}$ values are shown as black crosses in Figure \ref{fig:orbital energy}. The differences from the $a=1\,AU$ results are small (typically within 3.5\%). Moreover, these results are consistent with the earlier findings of \citep{2013ApJ...762...37L}, who adopted a semi-major axis of $\sim2.5\,AU$.

The cumulative effect over three encounters shows notable differences from the first passage behavior. 
While the energy transfer in the first encounter peaks at the pericenter distance around 1.4 $r_\mathrm{t}$, subsequent interactions shift this location of maximum net positive $  \Delta E_{\mathrm{orb}}  $  toward larger $r_\mathrm{p}/r_\mathrm{t}$ values. This trend is clearly visible in Figure~\ref{fig:orbital energy}, where the peak of cumulative $  \Delta E_{\mathrm{orb}}  $ after three passages lies at $  r_\mathrm{p} / r_\mathrm{t}  \sim 1.6 $.
This outward shift arises primarily because mass loss during the initial encounter preferentially removes low-density envelope material, thereby increasing the core mass ratio especially for deepest passages. A remnant with a higher core mass fraction becomes dynamically stiffer, exhibiting reduced tidal mass loss in subsequent evolution. As a result, the mass loss between the two tidal streams is more symmetric \citep{2013ApJ...762...37L}, and $  \Delta E_{\mathrm{orb}}  $ is dominated by the orbital energy dissipation by tides, which causes the orbit becomes more bound. This finding is of particular significance to the evolution of tidally disrupted giant planets, i.e., as long as giant planets survive from being ejected during the initial asymmetric mass loss phrase, we anticipate that subsequent periastron passages with more and more symmetric mass loss will circularize the planet's orbit. 
% A periastron passage that produced strong asymmetry in the first encounter may produce only mild asymmetry in the second, pushing the optimal asymmetry region outward. 

The $\Delta E_{\mathrm{orb}}$ for each successive encounter converges as the initial $r_\mathrm{p}/r_\mathrm{t}$ approaches 2.0. This convergence can be understood through the interplay of orbital parameters and structural evolution. Our results show that while the orbital semi-major axis $a$ can change significantly after an encounter, the change in the pericenter distance $r_\mathrm{p}$ itself is remarkably limited, on the order of one-tenth of a planetary radius due to the conservation of specific orbital angular momentum. For orbits with initial $r_\mathrm{p}/r_\mathrm{t} \sim 2$, each encounter results in minimal mass loss and consequently negligible structural change. Furthermore, the $r_\mathrm{p}$ values for the second and third encounters remain very similar. This combination—minimal evolution in both the planet's internal structure and the encounter geometry ($r_\mathrm{p}$)—leads to nearly identical energy exchanges in successive passages. 

\section{Discussion}
\label{sec:discussion}
\subsection{Survival and Orbital Evolution of Giant Planets with Dense Cores}
\label{sec:survival-and-orbit}
A central result of our simulations is that giant planets possessing dense cores are not subject to total tidal destruction during repeated close stellar encounters. In coreless models \citep{2011ApJ...732...74G}, a single deep passage could lead to catastrophic stripping, but the presence of even a small core ($10\,M_\oplus$) dramatically changes this outcome.
The dense core acts as a gravitational anchor, significantly increasing the binding energy of the surrounding envelope and suppressing large-scale deformation. As a consequence, even in the most destructive cases we simulated ($r_\mathrm{p}/r_\mathrm{t}=1.2$, and core mass = $10\,M_\oplus$), the planet loses substantial gaseous envelope mass after three encounters but a bound remnant stabilizes around $\sim 20\,M_\oplus$, comprising its pristine $10\,M_\oplus$ core and a comparable mass of residual gaseous envelope. 

Because the core's mean density ($\sim 20\,\mathrm{g}\,\mathrm{cm}^{-3}$) is an order of magnitude higher than that of the host star, it remains completely immune to tidal disruption. Here we stop the hydrodynamic simulations of mass stripping for a prolonged period of time as our fixed radius assumption is no longer a good approximation in the evolutionary regime dominated by the planetary core. This is previously studied with composite polytropes undergoing adiabatic mass loss \citep{2013ApJ...762...37L}. Once the envelope mass is sufficiently reduced, the size of the composite polytropes with a stiff dense core will shrink significantly causing mass loss to effectively cease. In reality, the residual gaseous envelope, loosely bound and extended, is highly vulnerable to continued mass loss. Processes such as ongoing tidal stripping during subsequent passages, photoevaporation driven by stellar irradiation, and thermal escape would likely to play a key role in the long-term evolution, which will be explored in future studies. 

The presence of a dense core not only prevents total tidal destruction but also significantly lowers the probability of planetary ejection during the high-eccentricity migration phase. In coreless models, deep encounters frequently produce strong asymmetric mass loss that imparts a large positive velocity kick to the remnant, often increasing its specific orbital energy enough to unbind the planet from the host star \citep{2011ApJ...732...74G}. By contrast, the gravitational anchoring effect of the core reduces the extent and asymmetry of envelope ejection per passage. This results in smaller positive (or even negative) $\Delta E_{\mathrm{orb}}$ values compared to coreless cases at the same periastron distance, making it substantially harder for the planet to achieve the escape velocity required for ejection after a single or several successive encounters.

However, whether a tidal encounter ultimately leads to ejection remains highly sensitive to the planet's initial specific orbital energy (i.e., the semi-major axis $a$ or eccentricity $e$). For planets starting on very wide, nearly parabolic orbits (large $a$), even a modest positive $\Delta E_{\mathrm{orb}}$ from asymmetric mass loss can push the total energy above zero, resulting in hyperbolic escape. Our choice of a reduced semi-major axis ($a = 1$ AU) was deliberately made to suppress this early-ejection pathway, allowing us to isolate the cumulative structural and orbital evolution over multiple passages. In more realistic dynamical excitation scenarios—where giant planets are scattered from several AU to high eccentricities—the initial $a$ is typically large enough that a single strong asymmetric mass-loss event could indeed eject the planet if the core were absent or very small. The stabilizing influence of even a $10\,M_\oplus$ core therefore acts in two complementary ways: (i) it curtails the magnitude of positive $\Delta E_{\mathrm{orb}}$ by limiting asymmetric stripping, and (ii) it preserves bound remnants that can continue to tighten their orbits through subsequent negative $\Delta E_{\mathrm{orb}}$ contributions from reduced asymmetry and eventual tidal dissipation.

Taken together, these results suggest that dense cores dramatically increase the survival probability of gas giants undergoing high-eccentricity migration—not only by preventing runaway disruption but also by reducing the likelihood of ejection during the critical early encounters. For planets that do survive the initial asymmetric phase, progressive envelope stripping and orbital circularization become the dominant evolutionary pathways, ultimately favoring the formation of compact, high-metallicity remnants rather than loss of the planet from the system.

Recent observations of ultra-dense planets like GJ 523b, classified as Mega-Earths \citep{kroft2026gj523bmassive170}, provide a potential observational counterpart to the core-dominated remnants that we predict. Its high density and lack of a substantial gaseous envelope are unexpected from standard core accretion models, while its high orbital obliquity suggests a dynamical history involving scattering or high eccentricity migration.

%\begin{figure*}
%    \centering
    %\includegraphics[width=0.95\textwidth]{ae_diagram_planets.pdf}  
    %\caption{Semi-major axis versus eccentricity diagram for high-eccentricity exoplanets. Overlaid are contours of constant tidal circularization time for 1 Gyr and 5 Gyr, computed for a Jupiter-mass planet with Jupiter's radius, for two representative values of the tidal dissipation parameter: $Q'_{\mathrm{p}} = 10^4$ (blue dotted line) and $Q'_{\mathrm{p}} = 10^6$ (black dash-dotted line). The inset panel at the bottom extends to $e\gtrsim0.7$ and overlays observed exoplanets with known masses, radii, and host-star ages (retrieved from the Extrasolar Planets Encyclopaedia (\url{http://exoplanet.eu/}) on 2025 December 28). Point sizes are proportional to planetary mass, and colors indicate stellar age in Gyr. Planets located near or within the circularization contours may achieve tidal circularization within the stellar age, while those far outside the contours are likely to remain stalled in their eccentric orbits.}
    %\label{fig:ae_diagram_planets}
%\end{figure*}
\begin{figure*}[t] 
    \centering
    \includegraphics[width=\textwidth]{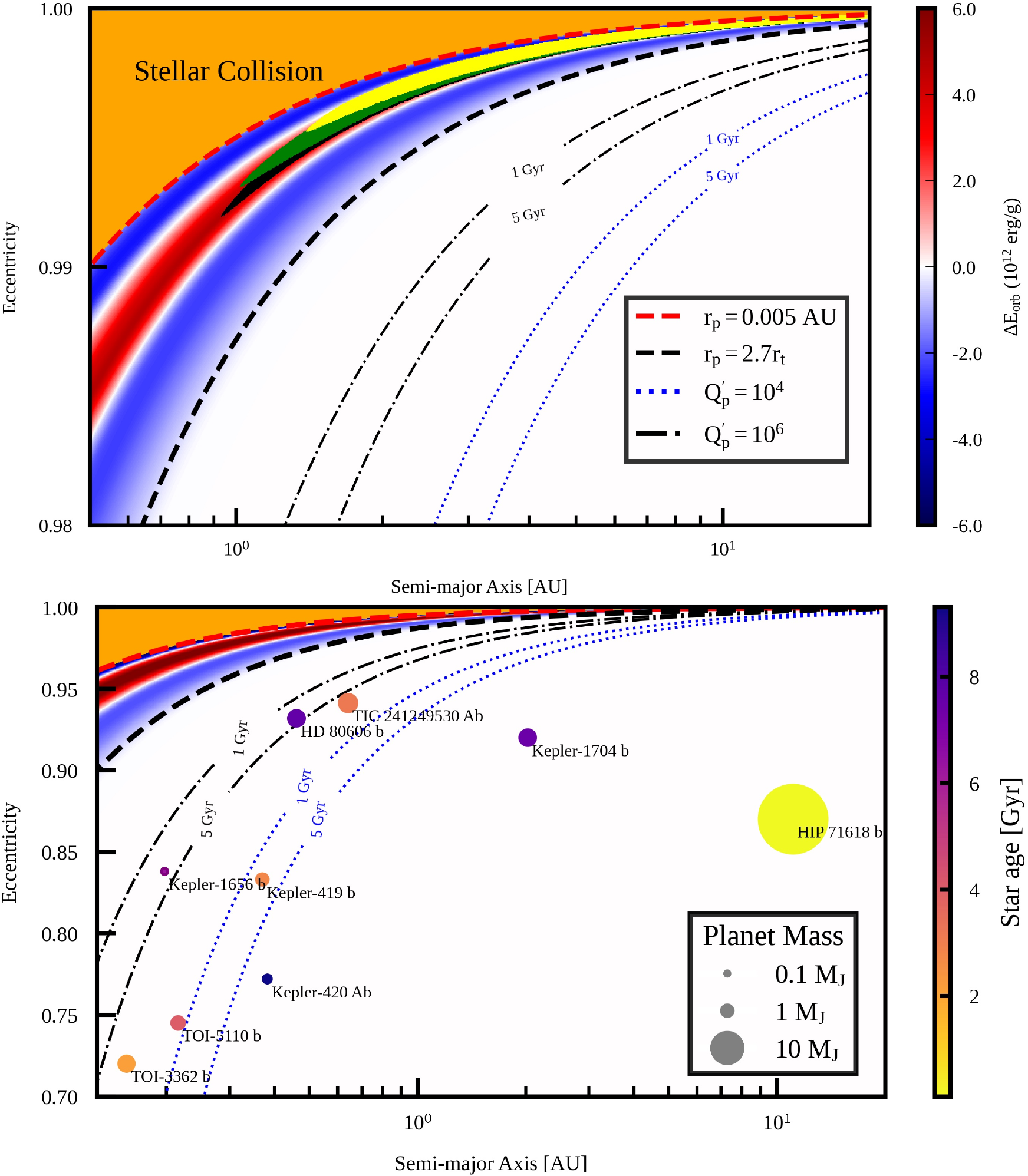} 
    \caption{Semi-major axis versus eccentricity diagram for high-eccentricity exoplanets, based on our simulation results for planets with a $10\,M_\oplus$ core. Overlaid are contours of constant tidal circularization time for 1 Gyr and 5 Gyr, computed for a Jupiter-mass planet with Jupiter's radius, for two representative values of the tidal dissipation parameter: $Q'_{\mathrm{p}} = 10^4$ (blue dotted line) and $Q'_{\mathrm{p}} = 10^6$ (black dash-dotted line). The lower panel extends to $e\gtrsim0.7$ and overlays observed exoplanets with known masses, radii, and host-star ages (retrieved from the Extrasolar Planets Encyclopaedia (\url{http://exoplanet.eu/}) on 2025 December 28). Point sizes are proportional to planetary mass, and colors indicate stellar age in Gyr. Planets located near or within the circularization contours may achieve tidal circularization within the stellar age, while those far outside the contours are likely to remain stalled in their eccentric orbits.}
    \label{fig:ae_diagram_planets}
\end{figure*}

\subsection{Generalization to a Broader a-e Parameter Space}
As demonstrated by additional simulations with varying semi-major axes (see Section~\ref{sec:Orbital Evolution}), the change in specific orbital energy $  \Delta E_{\mathrm{orb}}  $ for highly eccentric orbits depends primarily on the periastron distance $  r_\mathrm{p}  $ and only weakly on the semi-major axis $  a  $. This weak dependence on $a$ can be understood analytically from the small difference in tidal force between orbits with different $e$ at fixed $r_\mathrm{p}$.

\cite{2011ApJ...732...74G} quantified this effect by comparing the tidal force along two orbits with slightly different eccentricities. For a fixed $r_\mathrm{p}$, the relative difference in tidal force at the same true anomaly is given by
\begin{equation}
\frac{F_1 - F_2}{F_\mathrm{{r_\mathrm{p}}}} = r_\mathrm{p}^3 \left( r_1^{-3} - r_2^{-3} \right).
\end{equation}
Let \(\Delta F = F_1 - F_2\) denote the difference in tidal force between two orbits with eccentricities \(e_1\) and \(e_2\), and \(F_\mathrm{{r_\mathrm{p}}}\) the force at pericenter. Then,
\begin{equation}
    \frac{\Delta F}{F_\mathrm{{r_\mathrm{p}}}}\approx \frac{3(e_1 - e_2)}{e_2(e_2 + 1)} \frac{r_\mathrm{p}^3}{r_1^3} \left( \frac{r_1}{r_\mathrm{p}} - 1 \right),
\end{equation}
when $e_1 - e_2 \to 0$.
This expression is maximized at $r_1 = \frac{3}{2} r_\mathrm{p}$. Evaluating it for the extreme case $e_1 = 0.9$ and $e_2 \sim 1$ (i.e., a nearly parabolic orbit) yields a maximum force difference of $\sim 3.33\%$ at this location, confirming that the tidal perturbation near pericenter remains comparable.

%We also directly compare the orbital evolution for a fixed pericenter distance ($r_\mathrm{p}/r_\mathrm{t} = 1.2$, representing the most extreme mass loss) across three distinct orbital configurations: ($e=0.99$, $a\sim0.57\,\mathrm{AU}$), ($e\sim0.994$, $a=1\,\mathrm{AU}$), and ($e\sim0.999$, $a=5\,\mathrm{AU}$). In the left panel of Figure \ref{fig:orbital energy}, the black crosses indicate the simulation results where $a=5\,\mathrm{AU}$, with all other parameters held constant. The proximity of each cross to the red dot validates our approximation that $\Delta E_{\mathrm{orb}}$ depends only weakly on $a$, with a maximum deviation of only 3.5\%.These results confirm that the encounter outcome is indeed a robust function of $r_\mathrm{p}/r_\mathrm{t}$ for $\mathbf{e}\gtrsim0.9$.

This decoupling allows for a powerful generalization of our results obtained with a fixed initial $a=1\,\mathrm{AU}$.
We can map our discrete simulation results onto a continuous $\mathbf{a}-\mathbf{e}$ orbital parameter space. For any given ($\mathbf{a},\mathbf{e}$) pair, the corresponding $r_\mathrm{p} = a(1-e)$ and its ratio to the tidal radius $r_\mathrm{t}$ can be used to interpolate the expected $\Delta E_{\mathrm{orb}}$ per encounter from our calibrated relations. The result of this mapping is presented in Figure \ref{fig:ae_diagram} for core masses of $10\,M_\oplus$ and $20\,M_\oplus$. 

While the above scaling suggests that our results could in principle be extended to eccentricities as low as $e=0.9$, we restrict our subsequent generalization to $e\gtrsim0.99$ (i.e., when $e=0.99$, $a\approx0.57\,\mathrm{AU}$ for $r_\mathrm{p}/r_\mathrm{t}=1.2$ and $a\approx0.95\,\mathrm{AU}$ for $r_\mathrm{p}/r_\mathrm{t}=2.0$). This restriction ensures that the orbital period remains sufficiently long for the planet to relax back to a nearly spherical shape between encounters.

The color scale indicates the cumulative specific orbital energy change, $\Sigma\Delta E_{\mathrm{orb}}$, after three encounters, with red and blue hues denoting a net increase (orbital expansion) and decrease (decay), respectively. The red dashed line marks $r_\mathrm{p} = 0.005\, \mathrm{AU}$, slightly exterior to the solar radius. Orbits stirred into the region to its left would result in a stellar collision. Notice that a linear color scale is employed to best visualize the dynamic range of interest. In the lower-right region of the diagrams, where $r_\mathrm{p}$ is large, tidal dissipation dominates. The energy decay per passage there is several orders of magnitude smaller than the changes resolved in our disruption-focused simulations; hence, it appears white. The yellow, green, and black contours enclose orbits that would be ejected (i.e., the orbital energy becomes positive) after one, two, and three encounters, respectively. The black dashed line indicates $r_\mathrm{p}/r_\mathrm{t} \sim 2.7$. Previous work by \cite{2011ApJ...732...74G} (using a coreless polytropic model) suggested that planets stirred into the region $r_\mathrm{p}/r_\mathrm{t} \lesssim 2.7$ are either ejected or totally disrupted.

The red region, corresponding to moderate $r_\mathrm{p}/r_\mathrm{t}$ in our simulations, splits the blue region (energy decrease) into two distinct domains. In the innermost blue region (small $r_\mathrm{p}$), planets undergo rapid and severe mass loss, shedding nearly their entire envelopes within a few encounters. 
%After only a few passages, they are reduced to bare dense cores, effectively becoming rocky planets. These remnants remain on highly eccentric orbits but experience negligible further mass loss due to the core's structural rigidity. 

The red zone, where energy gain is positive, is the source of ejections. Planets that enter this region are readily unbound, potentially contributing to a population of free-floating planets (FFPs). The survivors, those with initially small semi-major axes or larger core masses, ultimately undergo a similar downsizing process, ending as rocky remnants.

The outer blue region (larger $r_\mathrm{p}$) presents a more complex picture. Its inner part, adjacent to the red zone, behaves analogously to the innermost blue region: planets eventually contract to rocky cores, but the timescale is longer. 
Farther out, where mass loss is minimal, tidal heating may gradually become important over long timescales. Whether this deposited energy can accumulate over many orbits to alter the planet's structure or orbital evolution is beyond the scope of this paper and will require coupling with thermal evolution models.
%Although each periastron passage deposits only a small amount of energy, the cumulative effect over many orbits could gradually inflate the planet's radius. An inflated envelope is more vulnerable to subsequent tidal stripping. Thus, such planets might experience modest mass loss despite their large $r_p$, ultimately becoming low-mass but extended objects. It's a potential pathway to the observed population of "puffy" planets. The efficiency of this mechanism depends sensitively on the internal dissipation and the orbital parameters, and we intend to investigate it in our future work. 
Beyond this heating-dominated regime, the long-term evolution is governed by tidal circularization rather than disruption. 

Accurately determining the circularization timescale for highly eccentric planets is challenging. Following \cite{1981A&A....99..126H}, \cite{1998ApJ...499..853E}, \cite{2008ApJ...678.1396J} and \cite{2009ApJ...698.1357J}, we estimate this timescale as
\begin{equation}
    \tau_{\mathrm{cir}} \sim \frac{1}{21} \frac{Q_{\mathrm{p}}'}{n} \frac{M_{\mathrm{p}}}{M_{\star}} \left( \frac{a}{R_{\mathrm{p}}} \right)^5 (1 - e^2)^{13/2},
\end{equation}
which should be regarded as an order-of-magnitude approximation. Accurate values would require numerical integration of the full tidal evolution equations. Note that this expression neglects the stellar tidal contribution; incorporating it would substantially complicate the formulation.
Figure \ref{fig:ae_diagram_planets} places our results in a broader observational context by overlaying tidal circularization timescales onto the $\mathbf{a}-\mathbf{e}$ diagram. The upper panel corresponds to the parameter space examined in Figure \ref{fig:ae_diagram}, where dynamical encounters dominate. Also shown are contours of constant circularization time (1 Gyr and 5 Gyr) for two representative values of the tidal dissipation parameter ($Q'_{\mathrm{p}} = 10^4$ and $10^6$), computed for a Jupiter-mass planet with Jupiter's radius. Colored dots in lower panel represent observed exoplanets with measured masses, radii, and host-star ages, with the color scale indicating stellar age in Gyr.

%Although the observed planets currently lie outside the dynamical encounter region shown in Figure \ref{fig:ae_diagram}, they are not necessarily primordial occupants of their current orbits. Many are likely in the process of tidal circularization, having originated from more eccentric orbits that may have once entered the dynamical encounter region. 
%For planets with moderately large pericenter distances (i.e., those falling within or near the circularization contours, like HD 80606 b), the tidal circularization timescale can be shorter than the stellar age; such planets will gradually circularize, ultimately settling into an orbit with semi-major axis $a_{\mathrm{final}} \sim 2r_{\mathrm{p}}$. In contrast, planets located far outside the contours, such as Kepler-1704 b and HIP 71618 b, have circularization timescales greatly exceeding their host star ages and are expected to remain essentially frozen in their current eccentric orbits. 
%For orbits with moderately large $r_\mathrm{p}$ (i.e., those lying in the white region closer to the colored zones), the tidal circularization timescale can be shorter than the stellar age; such planets will gradually circularize, ultimately settling into an orbit with semi-major axis $a_\mathrm{final}\sim 2r_\mathrm{p}$. Moving further outward, where $r_\mathrm{p}$ is so large that even Gyr of tidal evolution produce negligible change, planets remain essentially frozen in their original eccentric orbits. 

Although no observed exoplanets currently lie within or even close to the dynamical-encounter region ($  r_\mathrm{p}/r_\mathrm{t} \lesssim 2.7  $), this is expected. Meanwhile, only a handful of known giant planets have eccentricities $  e > 0.9  $. The scarcity arises because the closer a giant planet approaches the strong tidal zone, the shorter its orbital circularization timescale becomes.
For planets whose periastron distances lie near or inside the circularization contours, the circularization timescale is comparable to or shorter than the host-star age (depending on the planet's tidal quality factor $  Q'_\mathrm{p}  $). Such systems are therefore expected to be actively circularizing and will eventually settle onto a nearly circular orbit with final semi-major axis $  a_{\mathrm{final}} \sim 2 r_\mathrm{p}  $. Note that for super Jupiters, e.g., HD 80606 b with $  r_p \approx 0.031  $ AU and $  e \approx 0.932  $, the actual circularization timescale is likely longer due to higher mean densities. In contrast, planets such as Kepler-1704 b and HIP 71618 b have much further pericenters, resulting in circularization timescales that greatly exceed the stellar age. These planets are effectively frozen in their highly eccentric orbits on Gyr timescales.

\subsection{Existence of a Well-Defined Ejection Zone}

The contours separating surviving and ejected orbits in Figure~\ref{fig:ae_diagram} are remarkably sharp, revealing a well-defined ejection zone. This threshold-like behavior indicates that the transition from bound to unbound orbits is highly sensitive to the initial orbital parameters yet remains largely deterministic, governed by a critical change in specific orbital energy rather than stochastic processes.
%The structure of the ejection zone can be understood in terms of the orbital energy change $\Delta E_{\mathrm{orb}}$ per periastron passage. Ejection occurs when $\Delta E_{\mathrm{orb}}$ exceeds the current binding energy. Orbits that require two or three encounters lie just outside the immediate one-encounter ejection zone: the energy change per passage is insufficient to unbind the planet in a single event, but successive encounters deposit additional energy in small increments until the cumulative gain surpasses the binding energy. 

Ejection occurs when the cumulative change in specific orbital energy, $  \Sigma\Delta E_{\mathrm{orb}}  $, exceeds the planet's initial binding energy. Orbits that require two or three encounters to be ejected lie just outside the single-encounter ejection boundary: the energy gain per passage is insufficient to unbind the planet immediately, but successive encounters gradually accumulate enough positive $  \Delta E_{\mathrm{orb}}  $ to push the total energy above zero.

It is important to note that the ejection zone shown in Figure \ref{fig:ae_diagram} is derived from simulations limited to three encounters and therefore represents a strict lower bound on the region of parameter space from which planets can be ejected. Orbits lying just outside this zone are not necessarily safe over longer timescales; many would likely be ejected after additional encounters. We deliberately stopped at three encounters because, as discussed in Section \ref{sec:structure}, the fixed-radius assumption becomes increasingly inaccurate under the condition of extreme mass loss. Nevertheless, three encounters already constrain the majority of the ejection zone (relatively small $  r_\mathrm{p}/r_\mathrm{t} \sim 1.4  $--$1.7$) that separates the downsizing zone (smallest $  r_\mathrm{p}/r_\mathrm{t}  $) and the survival zone (intermediate and large $  r_\mathrm{p}/r_\mathrm{t}  $). We show in Figure \ref{fig:ae_diagram} that the first encounter contributes the largest fraction of the ejection region in the $  (a,e)  $ parameter space, while the second and third encounters add successively smaller fractions. Moreover, any additional ejection beyond three encounters would require an initial semi-major axis smaller than $  \sim 1  $ AU, which has a remote probability in HEM scenario. %The ejection zone is thus expected to expand gradually with increasing number of encounters before converging to a well-defined asymptotic boundary. 

This convergence of the tidal ejection zone arises from the structural evolution of the planet during mass loss. Giant planets with dense cores experiencing the strongest positive energy gains typically have intermediate periastron distances ($r_\mathrm{p}/r_\mathrm{t}\sim1.4-1.7$). After a few stripping encounters, these planets are reduced to a tenuous gas envelope surrounding a dominant dense core. The remnant becomes dynamically stiffer, which suppresses large-amplitude tidal deformation and reduces the magnitude of asymmetric mass loss in subsequent passages. As a result, the $  \Delta E_{\rm orb}  $ curves shift toward larger $  r_p/r_t  $ with successive encounters, the differences between the curves become progressively smaller ( as clearly visible in Figure \ref{fig:orbital energy}). Consequently, the cumulative energy gain approaches a limiting value, and the ejection zone ceases to expand further. 

Thus, the true boundary of the tidal ejection zone is ultimately set by the structural evolution of the planet during subsequent tidal encounters, i.e., more specifically, the transition to a core-dominated, stiffer configuration that limits further asymmetric mass loss. Our three-encounter ejection contours represent a conservative lower bound, and are sufficient to support the central claims of this paper, including the identification of distinct tidal regimes and the absence of total disruption for core-bearing planets. 

\section{Summary}
\label{sec:summary}

We have performed three-dimensional hydrodynamic simulations of the tidal evolution of core-bearing giant planets undergoing high-eccentricity migration. By following multiple successive periastron passages and mapping the results onto the continuous $  (a,e)  $ orbital parameter space, we demonstrate that even a modest dense core (as small as $10\,M_\oplus$) fundamentally alters the tidal outcome compared to coreless models.

Contrary to previous predictions that giant planets are completely disrupted inside $  r_\mathrm{p}/r_\mathrm{t} \approx 2.7  $, our simulations show no total disruptions when a dense core is present. Instead, assuming $a\gtrsim 1 $ AU \citep[i.e., the cold scattering defined in][]{2026arXiv260322409E}, three main outcomes emerge depending on the initial periastron distance:

\begin{itemize}
\item \textbf{Downsizing zone (smallest $  r_\mathrm{p}/r_\mathrm{t}  $):} At close pericenters, planets experience severe envelope stripping and are reduced to remnants consisting of the dense core plus a thin residual gaseous envelope. These objects are too low-mass to become hot Jupiters and are more likely to evolve into rocky super-Earths or hot Neptunes.
\item \textbf{Ejection zone (relatively small $  r_\mathrm{p}/r_\mathrm{t} \sim 1.4  $--$1.7$):} In this region, cumulative positive changes in specific orbital energy lead to ejection after one to three encounters. Such planets are removed from the system and may contribute to the free-floating planet population.
\item \textbf{Survival zone (intermediate and large $  r_\mathrm{p}/r_\mathrm{t}  $):} Planets with wider pericenters experience only minimal mass loss. These survivors can undergo gradual tidal circularization over Gyr timescales via tidal dissipation, eventually settling onto short-period orbits with final semi-major axis $  a_{\mathrm{final}} \sim 2\, r_\mathrm{p}  $, consistent with the formation of hot Jupiters.
\end{itemize}

These findings revise the classical picture of tidal disruption during high-eccentricity migration by demonstrating the critical stabilizing role of dense cores. While our simulations are limited to three encounters and employ a fixed-radius approximation, the qualitative trends are robust and highlight the importance of internal structure in determining planetary fate.

Future work will incorporate longer integration times, self-consistent radius evolution, and coupling between dynamical and thermal models to track tidal heating, envelope inflation, and the long-term evolution of partially stripped remnants.
Nevertheless, because the dense core remains largely unaffected by tidal heating and acts as a gravitational anchor, the absence of total disruptions reported here is expected to be robust.
Extending the parameter space to a broader range of core masses and initial orbital conditions will further clarify how high-eccentricity migration shapes the observed demographics of close-in giant planets.

Finally, this study considers only the direct star-planet tidal interaction on highly eccentric orbits. The influence of multi-planet dynamics—specifically whether close stellar encounters can destabilize the wider planetary system and trigger additional scattering events—remains an important direction for future investigation.

\begin{acknowledgments}
We are grateful to Yuhiko Aoyama, Dichang Chen, Fei Dai, Hongping Deng, Qiang Hou, Dong Lai, Doug Lin, Bo Ma, Zhizhen Qin, Xianyu Wang, Xing Wei, Cong Yu and Wei Zhong for helpful discussions. 
We also thank the anonymous referee for thoughtful suggestions which resulted in a greatly improved paper. This work is supported by the National Natural Science Foundation of China (grant Nos. 42530203, 11903089) and the Guangdong Basic and Applied Basic Research Foundation (grant No. 2021B1515020090). 
We acknowledge the School of Physics and Astronomy, Sun Yat-sen University, for providing computational resources on the Loong server. 
The software used in this work was developed in part by the DOE NNSA- and DOE Office of Science-supported Flash Center for Computational Science at the University of Chicago and the University of Rochester.
\end{acknowledgments}

\section*{APPENDIX}
\phantomsection
\label{app:profiles}

\begin{deluxetable*}{ccccc}
\tablecaption{\label{table:1}Model Parameters for Composite Polytropes}
\tablehead{
  \colhead{$M_\mathrm{P}$} & 
  \colhead{$R_\mathrm{P}$} & 
  \colhead{$\xi_\mathrm{1i}$} & 
  \colhead{$\xi_\mathrm{2i}$} & 
  \colhead{$\mu_1 / \mu_2$}
}
\startdata
  1.0 & 1.0 & 1.799 & 0.462 & 4.00 \\
  0.5 & 1.0 & 2.222 & 0.436 & 6.04 \\
  0.2 & 1.0 & 2.518 & 0.368 & 10.00 
\enddata
\tablecomments{All masses and radii are given in units of $M_\mathrm{J}$ and $R_\mathrm{J}$, respectively.}
\end{deluxetable*}

\begin{figure*}[t]
    \centering
    \includegraphics[width=\linewidth]{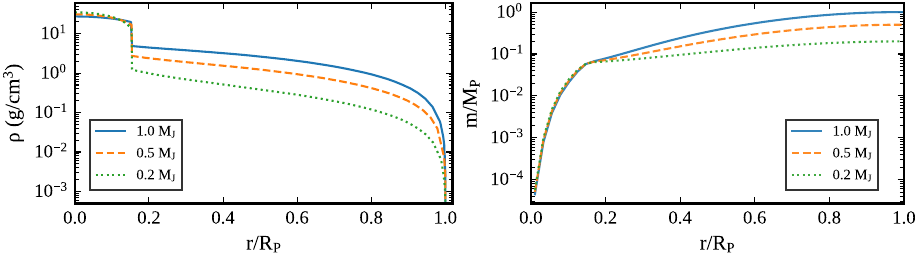} 
        \label{fig:sub1}
    \caption{Density (left) and cumulative mass (right) profiles for planetary models with a $20M_\oplus$ core, under different mass-loss scenarios: no mass loss, 50\% mass loss, and 80\% mass loss. The blue solid, orange dashed, and green dotted lines correspond to total masses of $1.0M_\mathrm{J}$, $0.5M_\mathrm{J}$, and $0.2M_\mathrm{J}$, respectively.}
    \label{fig:profiles}
\end{figure*}
Following \cite{2013ApJ...762...37L}, we model the core-envelope structure of a gas giant planet using a composite polytrope approach, accounting for distinct mean molecular weights in the core and envelope regions. With the polytropic assumption, $P$ is a simple power-law function of $\rho$
\begin{equation}
    P = K\rho ^{\gamma } = K\rho^{({1+\frac{1}{n} })},
\end{equation}
where K is a constant.
We choose the polytropic indices $n_{1}=0.5$ and $n_{2}=1$ in the core and envelope, respectively. Thus, in each part, we have 
\begin{equation}
 \rho_{1} = \rho_\mathrm{1c}\theta _{1}^{n_{1}},\ \rho_{2} = \rho_\mathrm{2i}\theta _{2}^{n_{2}},
 \end{equation}
and
\begin{equation}
R_\mathrm{core}=a_{1}\xi_\mathrm{1i} =a_{2}\xi_\mathrm{2i}\ , \ 
R_\mathrm{P}=a_{2}\xi_\mathrm{2s}.
\end{equation}
The subscripts c, i, and s denote the center, interface, and surface of the planet, respectively. Subscripts 1 and 2 refer to quantities associated with the core and the envelope, respectively. And $a_{1}$ and $a_{2}$ are given by the equation below:
\begin{equation}
    a_{1}=\sqrt{ \frac{K_{1}(n_{1}+1)}{4\pi G} \rho^{\frac{1-n_{1}}{n_{1}} }_\mathrm{1c}}=
\sqrt{\frac{3K_{1}}{8\pi G}\rho _\mathrm{1c} } ,
\end{equation}

\begin{equation}
    a_{2}=\sqrt{ \frac{K_{2}(n_{2}+1)}{4\pi G} \rho^{\frac{1-n_{2}}{n_{2}} }_\mathrm{2i}}=
\sqrt{\frac{K_{2}}{2\pi G} } .
\end{equation}
To obtain the entire structure of the planet, we have to solve the Lane-Emden equation of the core first:
\begin{equation}	
\frac{d^{2}\theta_{1} }{d\xi^{2}_{1}} +\frac{2}{\xi_{1}}\frac{d\theta_{1} }{d\xi_{1}} +\theta_{1} ^{n_1}=0.
\end{equation}
The boundary conditions are simple:
\begin{equation}
\theta _{1}(0)=1,
\end{equation}

\begin{equation}
\frac{d\theta_{1} }{d\xi_{1}}\Bigg |_{0}  =0.
\end{equation}

 To determine the interface of core and envelope, a cut-off point $\xi_\mathrm{1i}$ must be settled. At the interface, pressure and mass are continued:

\begin{equation}
    \frac{\xi _\mathrm{1i}\theta _\mathrm{1i}^{n_{1}}}{\theta _\mathrm{1i}^{'}\mu_{1}} =
\frac{\xi_\mathrm{2i}\theta _\mathrm{2i}^{n_{2}}}{\theta _\mathrm{2i}^{'}\mu_{2}} ,
\end{equation}

\begin{equation}
    (n_{1}+1)\frac{\xi_\mathrm{1i}\theta _\mathrm{1i}^{'}}{\theta _\mathrm{1i}\mu _{1}}=             (n_{2}+1)\frac{\xi_\mathrm{2i}\theta _\mathrm{2i}^{'}}{\theta _\mathrm{2i}\mu _{2}}.
\end{equation}

Here we set $\mu_{1}/\mu_{2}=4$. $\mu_1$ and $\mu_2$ denote the mean molecular weight in the core and envelope, respectively. The quantities of $\theta_\mathrm{2i}$ only affect $a_{2}$, so we take $\theta_\mathrm{2i}$ = 1 for simplicity. Thus, 
$\xi_\mathrm{2i}$ and $\theta_\mathrm{2i}^{'}$ can be calculated by
\begin{equation}
    \xi _\mathrm{2i} = \xi _\mathrm{1i}\frac{\mu _{2}}{\mu_{1}}\sqrt{\frac{n_{1}+1}{n_{2}+1}\frac{\theta _\mathrm{1i}^{n_{1}-1}}{\theta _\mathrm{2i}^{n_{2}-1}}}
=2\sqrt{3}\space \xi _\mathrm{1i}(\theta _\mathrm{1i})^{-\frac{1}{4} },
\label{eq:bc1}
\end{equation}
\begin{equation}
    \theta _\mathrm{2i}^{'}=\theta _\mathrm{1i}^{'}\frac{\mu_{2}}{\mu_{1}} \frac{\theta _\mathrm{2i}}{\theta _\mathrm{1i}} \frac{\xi_\mathrm{1i}}{\xi_\mathrm{2i}} \Big (\frac{n_{1}+1}{n_{2}+1}\Big)         
\theta _\mathrm{2i}^{'}=\frac{3}{16}  \frac{\theta _\mathrm{1i}^{'}}{\theta _\mathrm{1i}} \frac{\xi_\mathrm{1i}}{\xi_\mathrm{2i}}.
\label{eq:bc2}
\end{equation}
Equations \ref{eq:bc1} and \ref{eq:bc2} are also the boundary conditions in the envelope:
\begin{equation}
     \theta _{2}(\xi _\mathrm{2i})=1,
\end{equation}

\begin{equation}
    \frac{d\theta_{2} }{d\xi_{2}}\Bigg |_{\xi _\mathrm{2i}}=\frac{3}{16}  \frac{\theta _\mathrm{1i}^{'}}{\theta _\mathrm{1i}} \frac{\xi_\mathrm{1i}}{\xi_\mathrm{2i}}.
\end{equation}

By solving the Lane-Emden equation for the envelope and applying the boundary condition $\theta_{2}(\xi_\mathrm{2s})=0$, we determine the value of $\xi_\mathrm{2s}$ at the planetary surface. By adjusting the ratio $\mu_1/\mu_2$, we can construct planetary models with any desired total mass and radius while keeping the core mass and core radius fixed. Table \ref{table:1} lists the model parameters for several different masses.

% \begin{table}
% %\centering
% \caption{Model Parameters for Composite Polytropes}
% \label{table:1}
% \begin{tabular}{ c c c c c }
% \hline
% \hline
%  $M_\mathrm{P}(M_\mathrm{J})$  & $R_\mathrm{P}(R_\mathrm{J})$ & $\xi_\mathrm{1i}$ & $\xi_\mathrm{2i}$ & $\mu_1/\mu_2$\\
%  \hline
%  1.0  & 1.0 & 1.799 & 0.462 & 4.00 \\
%  0.5 & 1.0 & 2.222 & 0.436 & 6.04 \\
%  0.2 & 1.0 & 2.518 & 0.368 & 10.00 \\
 
%  \hline  
% \end{tabular}

% \end{table}

Figure \ref{fig:profiles} shows the density and cumulative mass profiles for models with a $20\,M_\oplus$ core under three mass-loss scenarios: no mass loss, 50\% mass loss, and 80\% mass loss, corresponding to total masses of $1.0$, $0.5$, and $0.2\,M_\mathrm{J}$, respectively. The blue solid, orange dashed, and green dotted lines represent these three cases. Although the central density of the core differs among models of varying total mass due to the nature of the polytropic model, they share a fixed core boundary radius and core mass. This consistency aligns with our assumption of a rigid, incompressible core throughout the tidal stripping process.
\bibliography{refs}{}
\bibliographystyle{aasjournal}


\end{CJK*}
\end{document}